\input{aipcheck}

\documentclass[
    ,final            
  ]
  {aipproc}

\layoutstyle{8x11double}

\begin{document}

\title{A loop quantum multiverse?}


\author{Martin Bojowald}{
  address={Institute for Gravitation and the Cosmos, The Pennsylvania State
    University, University Park, PA 16802, USA}
}

\begin{abstract}
  Inhomogeneous space-times in loop quantum cosmology have come under better
  control with recent advances in effective methods. Even highly inhomogeneous
  situations, for which multiverse scenarios provide extreme examples, can now
  be considered at least qualitatively.
\end{abstract}

\maketitle


\newcommand{\lP}{\ell_{\rm P}}
\newcommand{\md}{{\mathrm d}}
\newcommand{\tr}{{\mathrm{tr}}}
\newcommand{\vt}{\vartheta}
\newcommand{\vp}{\varphi}
\newcommand{\uvec}[1]{\raisebox{-1.5mm}{$\stackrel{\textstyle #1}{\scriptscriptstyle\rightarrow}$}{}}

\section{Introduction}

The question of whether there is one universe or a collection of different
ones in a multiverse is inherently inhomogeneous, and therefore requires any
quantum cosmological treatment to go beyond the common minisuperspace
constructions. It remains extremely difficult to address in theories such as
loop quantum gravity, which do not yet give rise to reliable intuitive and
tractable phenomena in anything but the simplest models. Nevertheless, recent
progress on effective descriptions of loop quantum gravity has revealed
general, perhaps even universal, effects at high curvature, which can be used
to test whether the theory makes it likely for the required structures of a
multiverse to form. Quite surprisingly, the cosmological scenarios based on
these new results are entirely unlike anything that has been imagined in most
homogeneous models of cosmology. Some quantum-geometry effects can be so
strong at high density that they trigger signature change, an implication
which has been overlooked for several years because it can only be seen when
inhomogeneity is implemented consistently. Consequences for a possible
multiverse are discussed in this article.

\section{Big-bang singularity}

In isotropic loop quantum cosmology \cite{LivRev,Springer}, the wave function
$\psi_{\mu}$, in terms of a geometrical variable $\mu$ quantizing the spatial
volume or the scale factor, can be extended to a universe before the big bang,
according to a difference equation of the form
\begin{equation} \label{Diff}
    C_+(\mu) \psi_{\mu+1}- 
C_0(\mu)\psi_{\mu}+
  C_-(\mu)\psi_{\mu-1}
  = \hat{H}_{{\rm matter}}(\mu)\psi_{\mu}\,.
\end{equation}
Following the recurrence, the wave function is evolved through $\mu=0$, the
classical singularity \cite{Sing,IsoCosmo}.  Departures not only from
classical dynamics but also from the continuous Wheeler--DeWitt equation arise
because there are strong ``holonomy modifications'' at nearly Planckian
density, to be discussed in more detail later in this article. These
corrections provide the terms by which the difference operator on the
left-hand side of (\ref{ModFried}) differs from a second-order derivative by
$\mu$ as it appears in the Wheeler--DeWitt equation. If finite shifts in the
difference operator are Taylor-expanded, a series of higher-order corrections
in the momentum of $\mu$ (a curvature component) is obtained
\cite{SemiClass}. Holonomy modifications therefore contribute to
higher-curvature corrections expected in any quantum theory of gravity, but
they lack higher time derivatives and therefore do not provide complete
curvature terms.

Motivated by the use of holonomies instead of connection components in the
full theory of loop quantum gravity, holonomy modifications replace the Hubble
parameter $H$ by a bounded function $\sin(\ell H)/\ell$ in a modified
Friedmann equation
\begin{equation} \label{ModFried}
 \frac{\sin(\ell H)^2}{\ell^2}= \frac{8\pi G}{3} \rho\,,
\end{equation}
with an ambiguity parameter $\ell$ of the dimension of length (possibly
related to the Planck length).  Using this modified Friedmann equation, one
obtains in simple models an effective picture of singularity resolution given
by a bounce \cite{APS}: Clearly, the energy density for solutions to
(\ref{ModFried}) must always remain bounded.

Note that higher-order corrections in an expansion of $\sin(\ell
H)^2/\ell^2\sim H^2(1+O(\ell^2 H^2))$ are indeed of the same size as usually
expected for higher-curvature corrections, given by the matter density divided
by something close to the Planck density. However, (\ref{ModFried}) ignores
higher time derivatives which should be of similar size as they, too,
contribute to higher curvature terms. As long as these terms are ignored, one
cannot use (\ref{ModFried}) at high density, near the maximum of the sine
function, and it remains unclear whether loop quantum cosmology generically
gives rise to bounces as an effective picture of its singularity
resolution. In some models with specific matter content (a free massless
scalar or kinetic domination), one can show that higher time derivatives are
absent or small \cite{BouncePert}. There is therefore a class of models in
loop quantum cosmology in which the mechanism of singularity resolution can
effectively be described as a bounce, and we will explore this scenario in
more detail here. Toward the end of the article, we will comment on the
entropy problem \cite{Tolman} which must be addressed in any bounce model.

A bounce, in general terms, may give rise to a multiverse picture, using the
following line of arguments: Consider a collapsing (part of the)
universe. Inhomogeneity builds up as the universe evolves.  If space is viewed
as a patchwork of nearly homogeneous pieces, their size must be decreased
after some time interval to maintain a good approximation, a mathematical
process which can be seen as describing the dynamical fractionalization of
space.  These statements are true in any collapsing universe. If one uses a
theory that gives rise to a bounce mechanism, denser patches that reach
Planckian density earlier bounce first (assuming that homogeneous models are
good for the patch evolution).  These bounced patches appear as expanding
regions embedded within a still-contracting neighborhood. Given the opposite
expansion behaviors, it is difficult to imagine that they can maintain causal
contact with their neighborhood. A multiverse picture not unlike that of
bubble nucleation in inflation results, except that there is no analog of
``bubbles within bubbles'' because a patch, once it has bounced, keeps
expanding and diluting for a long time and would have to recollapse to trigger
new bounces.

\begin{figure}
  \includegraphics[height=.2\textheight]{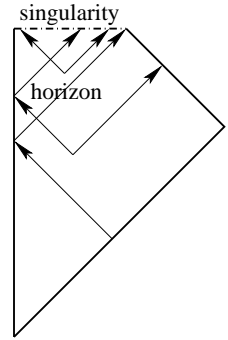}
  \caption{A classical space-time diagram for black-hole collapse, culminating
    in a singularity covered by a horizon. If space-time is spherically
    symmetric, the vacuum part of the interior within the horizon takes on the
    form of a homogeneous cosmological model. (This figure as well as
    Figs.~\ref{Interior} and \ref{Branch} are taken from
    \cite{Once}.) \label{BlackHoleEng}}
\end{figure}

\begin{figure}
  \includegraphics[height=.4\textheight]{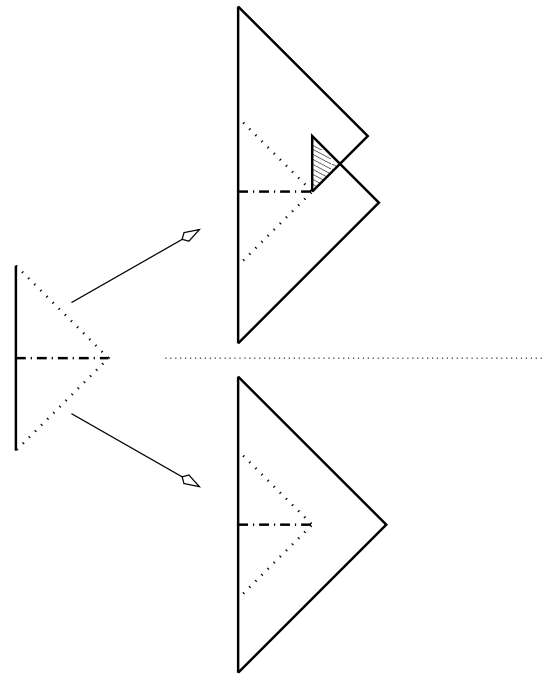}
  \caption{The vacuum interior region of Fig.~\ref{BlackHoleEng} can be
    quantized with methods of loop quantum cosmology, removing the classical
    singularity. The enlarged, non-singular interior may be embedded in
    spherically symmetric space-time in two causally different ways. (It is
    not known at present how precisely to construct such embeddings.) First,
    the post-singularity interior may open up into a new space-time without
    causal contact with the original outside region. Secondly, the interior
    may re-open into the original space-time, spilling out its matter as some
    kind of black-hole explosion. \label{Interior}}
\end{figure}

\begin{figure}
  \includegraphics[height=.25\textheight]{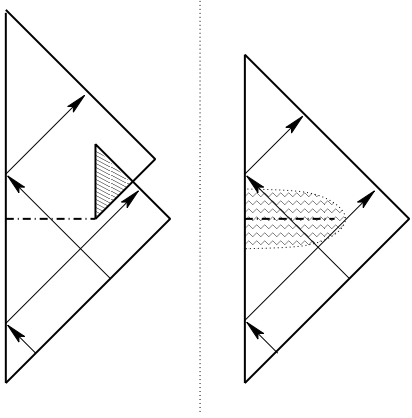}
  \caption{The two alternatives shown in Fig.~\ref{Interior} give rise to
    different space-time models. A baby universe is obtained from every black
    hole if the post-singularity interior does not connect causally back to
    the original space-time. Many such processes would then give rise to a
    multiverse. If the interior reconnects to the original space-time, black
    holes are merely compact, extremely dense objects within a single
    universe. \label{Branch}}
\end{figure}

Another, independent mechanism that, too, starts with general properties of
inhomogeneous collapse uses features of black holes: As inhomogeneity
increases, black holes may form and grow. If the singularities they contain
classically are resolved in quantum gravity, the question is where this dense
space-time region leads to. The interior space-time within the horizon of a
Schwarzschild black hole, Fig.~\ref{BlackHoleEng}, can be treated like
cosmological models and is resolved just like the big-bang singularity if
modifications of loop quantum cosmology are used \cite{BHInt}. Also here, the
presence of largely uncomputed higher-curvature corrections means that no
effective space-time for a non-singular black hole is known. But one can try
to see how a non-singular, possibly bouncing interior could be embedded within
an inhomogeneous black-hole space-time. The non-singular interior may be
embedded in a spherically symmetric exterior in different ways, shown in
Fig.~\ref{Interior}. If space-time splits off into a baby universe, a causally
disconnected region is obtained, as illustrated in Fig.~\ref{Branch}. Multiple
such processes provide a multiverse.

We need good control on inhomogeneity if we want to fill these scenarios with
more details. This task is difficult to achieve in non-perturbative quantum
gravity, but we can use effective theory to include the key effects in a
tractable model.

\section{Space-time structure}

\begin{figure}
  \includegraphics[height=.11\textheight]{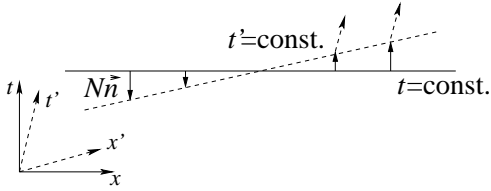}
  \caption{Lorentz boosts can be viewed as linear deformations of spatial
    slices in Minkowski space-time. Here, the standard Minkowski diagram is
    redrawn with a slightly different viewpoint, focussing on equal-time
    spatial slices in space-time. Tilted axes show how orthogonality in
    Minkowski geometry is to be represented after a boost to
    $(t',x')$. \label{LorentzMink}}
\end{figure}

In order to develop ingredients for an effective description of quantum
space-time in loop quantum gravity, we begin with a formulation and
generalization of the relevant classical structures. Classical space-time has
as symmetries Poincar\'e transformations, which one may view as linear
deformations of spatial slices in space-time: we have deformations along the
normal $\vec{n}$ by
\[
 N(\vec{x})=c\Delta
    t+\frac{\vec{v}\cdot \vec{x}}{c}
\]
with time translations and boosts, see Fig.~\ref{LorentzMink}, or within a
slice along $\vec{w}(\vec{x})= \Delta \vec{x}+{\bf R}\vec{x}$ with spatial
translations and rotations.

\begin{figure}
  \includegraphics[height=.2\textheight]{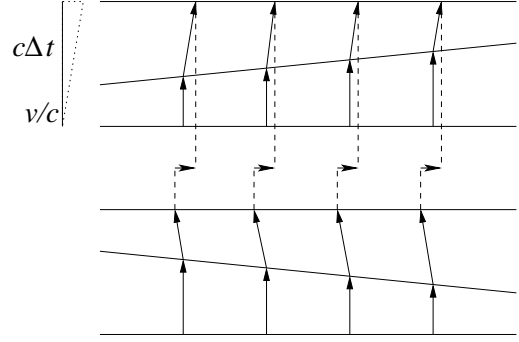}
  \caption{The Poincar\'e algebra follows geometrically from combinations of
    deformations as in Fig.~\ref{LorentzMink}. Shown here is the example of a
    (boost, time-translation) commutator, with the two orderings shown at the
    top and bottom. A normal deformation by $N_1(x)=v x/c$ (Lorentz boost) and
    one by $N_2(x)=c\Delta t-v x/c$ (reverse Lorentz boost and waiting
    $\Delta t$) commute up to a spatial displacement $w(x)=\Delta x=v\Delta
    t$, as computed using the small triangle with angle
    $v/c$. \label{HypDefLinMink}}
\end{figure}

\begin{figure}
  \includegraphics[height=.12\textheight]{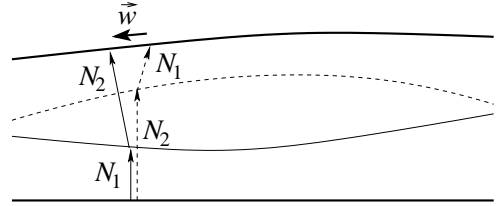}
  \caption{The space-time structure of general relativity is obtained by
    allowing for non-linear deformations of spatial slices. The Poincar\'e
    algebra is replaced by the infinite-dimensional hypersurface-deformation
    algebra. Using classical space-time geometry, one can derive
    (\ref{DD})--(\ref{HH}); see e.g.\ \cite{Action}. \label{SurfaceDefMink}}
\end{figure}

Algebraic calculations of commutators can be replaced by geometrical pictures,
such as the one shown in Fig.~\ref{HypDefLinMink}. In this way, it turns out,
one is more open to potential modifications of the algebra due to quantum
effects. It is also possible to extend the picture to general relativity
without much effort. We simply view local Lorentz transformations or
non-linear coordinate changes as non-linear deformations of space, as in
Fig.~\ref{SurfaceDefMink}.  Instead of the well-known commutators of the
Poincar\'e algebra, we obtain the not-so-well-known hypersurface-deformation
algebra of infinitely many generators $(S(\vec{w}(\vec{x})), T(N(\vec{x})))$,
labeled by a vector field $\vec{w}(\vec{x})$ and a function $N(\vec{x})$ in
space, with \cite{DiracHamGR}
\begin{eqnarray}
 [S(\vec{w}_1),S(\vec{w}_2)]&=& S((\vec{w}_2\cdot\vec{\nabla})\vec{w}_1-
 (\vec{w}_1\cdot\vec{\nabla})\vec{w}_2) \label{DD}\\
{} [T(N),S(\vec{w})] &=& T(\vec{w}\cdot\vec{\nabla}N)\\
{} [T(N_1),T(N_2)] &=& S(N_1\vec{\nabla}N_2-N_2\vec{\nabla}N_1)\,. \label{HH}
\end{eqnarray}

Hypersurface deformations not only generalize the Poincar\'e algebra, they
also geometrize the dynamics of space-time.  As shown by
\cite{Regained,LagrangianRegained}, second-order field equations for metrics,
invariant under the hypersurface-deformation algebra, must equal
Einstein's. Moreover, invariance under the hypersurface-deformation algebra
implies general covariance.  

All this is classical physics.  The problem of quantum gravity can be
approached by asking: How does quantum physics change hypersurface
deformations?

\section{Canonical gravity}

In order to see how quantum gravity may affect the relations
(\ref{DD})--(\ref{HH}), we must find operators that quantize the classical
expressions of $T(N)$ and $S(\vec{w})$. Posing the question in this way
suggests that canonical quantum gravity might be closests to answering it, and
loop quantum gravity is currently the best-developed canonical approach. In
this framework, computing the quantum version especially of (\ref{HH}) in
complete detail remains challenging, but a diverse set of methods, including
but not restricted to effective techniques, has during the last few years led
to mutually consistent and apparently universal results in most of the model
systems usually considered in general relativity
\cite{ConstraintAlgebra,LTBII,JR,ThreeDeform,ScalarHol,ModCollapse,TwoPlusOneDef,TwoPlusOneDef2,AnoFreeWeak,AnoFreeWeakDiff};
see \cite{Action,ReviewEff} for a general discussion.

Since we will be using methods of loop quantum gravity \cite{Rov,ThomasRev},
we describe space-time geometry by a canonical pair of an su(2)-valued
``electric field'' $\vec{E}_i$ and a ``vector potential'' $\uvec{A}_i$. (An
under-arrow indicates a covariant vector, or a 1-form.) The gravitational
electric field is a triad and determines spatial distances and angles by three
orthonormal vectors $\vec{E}_i$, $i=1,2,3$, at each point in space. The
gravitational vector potential $\uvec{A}_i$ is a combination of different
measures of curvature of space: the Ashtekar--Barbero connection.

In quantum field theory, one uses integrated (smeared) fields to construct
creation operators by which one can generate all Fock states out of the
vacuum.  In loop quantum gravity, the geometrical fields offer a natural
smearing of $\uvec{A}_i$ along curves (exponentiated to holonomies) and
$\vec{E}_i$ over surfaces (fluxes). Loop quantum gravity uses holonomies as
creation operators to construct a state space.  In what follows, we illustrate
these objects using a U(1)-connection $\uvec{A}$ for simplicity.  Holonomies
are then $h_e=\exp(i\int_e\md \lambda \uvec{A}\cdot\vec{t}_e)$ integrated
along curves $e$ in space, with tangent $\vec{t}_e$. For every possible $e$,
$\hat{h}_e$ provides excitations of geometry along this curve: As we will see
momentarily, surfaces intersected by $e$ gain area as the excitation level on
$e$ is increased.

To construct the corresponding quantum theory, we start
with a basic state $\psi_0$, $\psi_0(\uvec{A})=1$.  Excited states are
obtained by acting with holonomies:
\begin{eqnarray}
 \psi_{e_1,k_1;\ldots;e_i,k_i}(\uvec{A})&=&
\hat{h}_{e_1}^{k_1}\cdots \hat{h}_{e_i}^{k_i}\psi_0(\uvec{A})\\
 &=&\prod_{e}
h_e(\uvec{A})^{k_e}=\prod_{e} \exp(ik_e \smallint_e \md \lambda 
\uvec{A}\cdot\vec{t}_e)\,,\nonumber
\end{eqnarray}
written in the connection representation.  Many excitations along edges are
needed to obtain a macroscopic, near-continuum space-time region. Quantum
space-time is realized when only a small number of curves $e$ is geometrically
excited, and the action of a single holonomy operator has strong implications
on the overall state. Holonomy corrections to the classical equations, as
already used in (\ref{ModFried}), are then significant.

Derivative operators are quantized fluxes $\int_S\md^2y
\uvec{n}\cdot\hat{\!\vec{E}}$ for surfaces $S$ in space, with co-normal
$\uvec{n}$. They act by
\begin{eqnarray} \label{Flux}
  \int_S\md^2y \uvec{n}\cdot\hat{\!\vec{E}}\psi_{g,k}&=&
\frac{8\pi G\hbar}{i}\int_S\md^2y \uvec{n}\cdot \frac{\delta
  \psi_{g,k}}{\delta \uvec{A}(y)}\\
&=& 
8\pi \ell_{\rm Pl}^2\sum_{e\in g} k_e {\rm  Int}(S,e) \psi_{g,k}\nonumber
\end{eqnarray}
with the intersection number ${\rm Int}(S,e)$ and the Planck length $\ell_{\rm
  Pl}=\sqrt{G\hbar}$. On the right-hand side, we sum only integers, implying a
discrete spectrum for fluxes. The same kind of discreteness is realized if we
go back to SU(2)-valued fields, in which case we simply replace derivatives by
angular-momentum operators, and integers $k_e$ by spin quantum numbers
\cite{RS:Spinnet}. Geometry is discrete: for gravity, fluxes with discrete
spectra represent the spatial metric.

\subsection{Dynamics}

For the discrete dynamics of cosmic expansion, one must quantize the
gravitational Hamiltonian (constraint). Since it depends on the connection but
only holonomies can be represented as operators in loop quantum gravity,
modifications as in (\ref{ModFried}) are necessary, but now for the full
theory \cite{RS:Ham,QSDI}. The modified dynamics then takes into account
details of how discrete space grows by creating new lattice sites (atoms of
space), changing the excitation level of geometry as measured by fluxes.

The classical form of the Hamiltonian is somewhat analogous to that of
Yang--Mills theory on Minkowski space-time, where
\begin{equation} \label{HYM}
H=\kappa \int{\rm d}^3x
(|\vec{E}_i|^2+|\vec{B}_i|^2)
\end{equation}
for $\vec{B}_i= \uvec{\nabla}\times\uvec{A}_i+ C_{ijk}\uvec{A}_j\times
\uvec{A}_k$ (and structure constants $C_{ijk}$). For gravity on any
space-time, only showing the crucial terms,
\begin{equation} \label{H}
H(N)=\frac{1}{16\pi G}\int{\rm d}^3x
N\frac{\sum_{ijk}\epsilon_{ijk}(\vec{B}_i\times
  \vec{E}_j)\cdot\vec{E}_k}{\sqrt{\frac{1}{6}|\sum_{ijk}\epsilon_{ijk}
    (\vec{E}_i\times \vec{E}_j)\cdot\vec{E}_k|}} +\cdots
\end{equation}
with $C_{ijk}=\epsilon_{ijk}$. The presence of a free function $N$ in
(\ref{H}), as opposed to (\ref{HYM}), realizes the freedom of one's choice of
time coordinate in generally covariant theories. Indeed, $H(N)$, plus matter
Hamiltonians, plays the role of the time deformation generator $T(N)$
introduced earlier. If we can quantize $H(N)$ and compute commutators of the
resulting operators, we can see if and how (\ref{HH}) and the space-time
structure it encodes might change.  Not surprisingly, the required
calculations are rather complicated and remain incomplete, but some results
are known.

The form of the Hamiltonian together with properties of the loop
representation implies characteristic corrections when quantized.  We have
already mentioned higher-order corrections resulting from an expansion of
holonomies by $\uvec{A}_i$, used in place of the classical $\vec{B}_i$ in
(\ref{H}). Holonomy corrections will modify any gravitational Hamiltonian in
loop quantum gravity, and therefore indicate that (\ref{HH}) might be quantum
corrected. However, holonomy corrections are not the only ones. There is also
quantum back-reaction, which is present in any interacting quantum theory and
gives rise to higher-time derivatives and curvature corrections. In canonical
quantizations, effective techniques provide systematic methods to compute such
terms \cite{EffAc,Karpacz,HigherTime}. Note that holonomy corrections and
quantum back-reaction (higher-time derivatives) both depend on the curvature.
The magnitude of holonomy corrections and quantum back-reaction therefore
cannot easily be distinguished, but their algebraic features are sufficiently
different from each other to disentangle their implications for quantum
space-time, using the substitutes of (\ref{HH}) they imply.

\begin{figure}
  \includegraphics[height=.23\textheight]{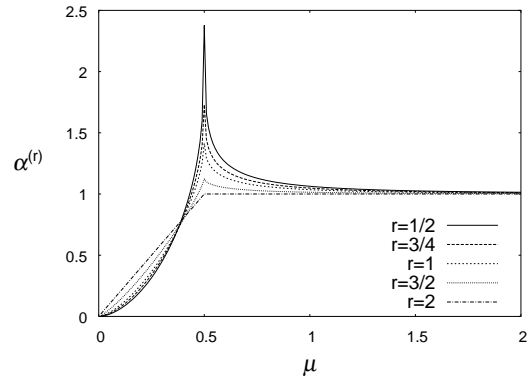}
  \caption{Inverse-triad corrections of loop quantum gravity imply a
    correction function $\alpha(\mu)$ that depends on ratios $\mu$ of flux
    eigenvalues by the Planck area and multiplies any occurrence of inverse
    triads for instance in Hamiltonians. These functions approach the
    classical limit $\alpha=1$ from above for large fluxes (coarser spatial
    lattices) and cut off classical divergences by strong quantum corrections
    at small flux values. The parameter $r$ describes a quantization
    ambiguity, which does not affect qualitative features such as $\alpha$
    becoming small near $\mu=0$ and $\alpha$ approaching one from above for
    $\mu\to\infty$.  \label{alpha}}
\end{figure}

There is a third type of corrections in loop quantum gravity which is easier
to handle and which we will discuss first.  We obtain inverse-triad
corrections from quantizing \cite{QSDV}
\begin{equation} \label{Inv}
 \left\{\uvec{A}^i,\int{\sqrt{|\det E|}}\mathrm{d}^3x\right\}= 2\pi G
 \epsilon^{ijk} \frac{\vec{E}_j\times\vec{E}_k}{{\sqrt{|\det E|}}}
\end{equation}
whose right-hand side appears in the Hamiltonian (\ref{H}) but cannot be
quantized directly, owing to non-existing inverses of flux operators
(\ref{Flux}) with discrete spectra containing zero. The left-hand side of
(\ref{Inv}), on the other hand, can be quantized and is regular, but implies
quantum corrections especially for small flux eigenvalues;
Fig.~\ref{alpha}. These corrections can be computed explicitly in models and
provide an automatic cut-off of the $1/E$-divergences
\cite{InvScale,QuantCorrPert,InflTest}. They refer to flux eigenvalues in
relation to the Planck scale, and are therefore independent of holonomy and
higher-curvature corrections which depend on the curvature scale or the energy
density. Inverse-triad corrections are therefore more reliable than holonomy
corrections, given that higher-curvature terms in loop quantum gravity remain
largely unknown.

\subsection{Inverse-triad corrections}

For any type of corrections, we can study dynamical implications by inserting
their correction functions in the classical Hamiltonian. For inverse-triad
corrections, for instance, we have
\begin{equation}
 \frac{1}{16\pi G}\int{\rm
  d}^3xN \alpha(\vec{E}_l) \frac{\sum_{ijk}\epsilon_{ijk}(\vec{B}_i\times
  \vec{E}_j)\cdot\vec{E}_k}{\sqrt{\frac{1}{6}|\sum_{ijk}\epsilon_{ijk}
(\vec{E}_i\times
    \vec{E}_j)\cdot\vec{E}_k|}} +\cdots
\end{equation}
with a correction function $\alpha(\vec{E}_l)$ as in Fig.~\ref{alpha}.  The
Hamiltonian generates time translations as part of the
hypersurface-deformation algebra; when the Hamiltonian is modified, the
Poisson-bracket algebra is therefore different from the classical one,
(\ref{DD})--(\ref{HH}). By consistency conditions of gravity as a gauge
theory, quantum corrections deform but do not violate covariance
\cite{ConstraintAlgebra}. We have
\begin{eqnarray}
 [S(\vec{w}_1),S(\vec{w}_2)]&=& S((\vec{w}_2\cdot\vec{\nabla})\vec{w}_1-
 (\vec{w}_1\cdot\vec{\nabla})\vec{w}_2) \label{DDalpha}\\
{} [T(N),S(\vec{w})] &=& T(\vec{w}\cdot\vec{\nabla}N)\\
{} [T(N_1),T(N_2)] &=&
S(\alpha^2(N_1\vec{\nabla}N_2-N_2\vec{\nabla}N_1))\,. \label{HHalpha} 
\end{eqnarray}
The algebra of hypersurface deformations is deformed, and the laws of motion
on quantum space-time change.

\begin{figure}
  \includegraphics[height=.2\textheight]{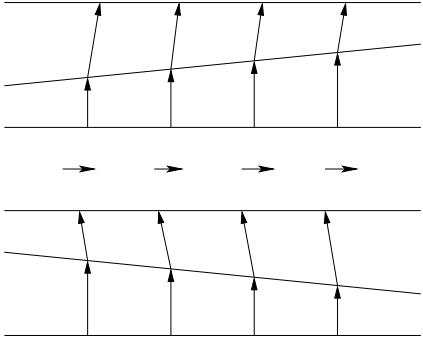}
  \caption{The hypersurface-deformation algebra (\ref{HHalpha}) in the
    presence of inverse-triad corrections modifies the classical laws of
    motion. With the constructions in Fig.~\ref{HypDefLinMink}, spatial
    displacements during a time $\Delta t$ at velocity $v$ differ from the
    classical value: $\Delta x=\alpha^2 v\Delta t$. Classical geometry can no
    longer be used to derive this relation, but it can be computed from the
    commutator (\ref{HHalpha}) with the functions $N_1$ and $N_2$ given in
    Fig.~\ref{HypDefLinMink}. Since $\alpha>1$ in mildly-modified,
    semiclassical regimes, the displacement is larger than expected
    classically and velocities seem to increase compared to the expected
    $v$. Even the speed of light is larger than classically, but there is
    still a consistent causal structure \cite{Tensor} thanks to the closed
    algebra (\ref{HHalpha}).  \label{HypDefLinDefMink}}
\end{figure}

Repeating the construction in Fig.~\ref{HypDefLinMink}, but using the deformed
algebra as in Fig.~\ref{HypDefLinDefMink}, the relation of a spatial
displacement to the boost velocity is modified: $\Delta x=\alpha^2v\Delta
t$. Discrete space speeds up propagation. (According to Fig.~\ref{alpha}, we
have $\alpha>1$ unless we are in strong quantum regimes of small $\mu$ in
which additional corrections have to be taken into account.)

Modified velocities can also be seen in cosmological perturbation equations,
in which the dynamics of density perturbations $u$ and gravitational waves $w$
now is \cite{LoopMuk}
\begin{eqnarray}
 -u''+s(\alpha)^2\Delta u +(\tilde{z}''/\tilde{z})u&=&0\,, \label{u}\\
 -w''+\alpha^2\Delta w +(\tilde{a}''/\tilde{a})w&=&0\,. \label{w}
\end{eqnarray}
The function $\alpha$ that modifies the speed of gravitational waves is shown
in Fig.~\ref{alpha}, and all other quantum-corrected functions, $s(\alpha)$,
$\tilde{a}$ and $\tilde{z}$ are known in terms of $\alpha$ but by rather
involved equations. As one general consequence, $s(\alpha)\not=\alpha$.  We
therefore have different speeds for different modes, and obtain corresponding
corrections to the tensor-to-scalar ratio as a characteristic signature of
deformed space-time.  For falsifiability of the theory, it is crucial that
$\alpha-1$ becomes large for small lattice spacing (flux values), while large
lattice spacing implies discretization effects and strong violations of
continuum physics.  Two-sided bounds on the discreteness scale are obtained,
improving the common dimensional expectations (effects of the size of the
average density divided by the Planck density) by several orders of magnitude
\cite{InflConsist,InflTest}.

Before we move on to the other types of quantum corrections, we discuss the
conceptual nature of quantum space-time subject to
(\ref{DDalpha})--(\ref{HHalpha}) or related modifications. No effective line
element can exist, for the metric and coordinate differentials transform in
non-matching ways, the metric --- a phase-space function --- according to
modified gauge transformations by (\ref{HHalpha}), but any ${\rm d}x$ by
standard coordinate changes. One could try to find new differential-geometry
structures, such as non-commutative \cite{Connes,NonCommST} or fractal ones
\cite{Fractional}, that could provide an invariant line element together with
metric coefficients subject to modified gauge transformations. But even
without a concrete space-time model, quantum space-time is well-defined
because all observables can be computed from (\ref{DDalpha})--(\ref{HHalpha})
by canonical methods. (See \cite{CUP} for canonical gravity.)  Quantum
space-time is also covariant since the full gauge algebra is realized, albeit
in a deformed manner \cite{DeformedRel}. Quantum space-time is just not
Riemannian space-time, but this is not to be expected anyway given the
presence of discrete structures.

\subsection{Quantum back-reaction}

Higher-curvature terms in effective quantum gravity
\cite{BurgessLivRev,EffectiveGR} modify the classical dynamics, but, unlike
quantum-geometry corrections of loop quantum gravity, leave the algebra
(\ref{HH}) unchanged \cite{HigherCurvHam}: they give theories in which the
space-time structure is still classical. They can be derived by standard
methods of low-energy effective actions, in which a derivative expansion
expresses non-locality by higher time derivatives. Covariance of the theory
together with a Poincar\'e-invariant vacuum state, around which the low-energy
effective action expands the quantum dynamics, imply that corrections can only
be by space-time scalars, or curvature invariants of increasing order in the
derivative expansion.

Loop quantum gravity, as a canonical theory, does not allow easy applications
of standard low-energy effective actions which are often based on path
integrals. Moreover, it is not clear whether it has a Poincar\'e-invariant
vacuum (or other) state, or whether such a state would be the right base to
expand around for, say, quantum cosmological phenomena. It is therefore
necessary to use a procedure which is canonical, and at the same time general
enough to encompass different quantum states. 

Such a procedure \cite{EffAc,Karpacz} can be found by turning Ehrenfest's
equations into systematic expansions. Ehrenfest's equations in quantum
mechanics express the rates of change of expectation values of basic operators
in terms of other, usually more complicated expectation values. For instance,
in quantum mechanics of a particle of mass $m$ in a potential $V(x)$, we have
\begin{equation} \label{x}
  \frac{{\rm d}\langle\hat{x}\rangle}{{\rm d}t}=
  \frac{\langle[\hat{x},\hat{H}]\rangle}{i\hbar}=
  \frac{\langle\hat{p}\rangle}{m}
\end{equation}
of the simple classical form. The expectation value of the momentum, however,
changes to
\begin{equation} \label{p}
 \frac{{\rm d}\langle\hat{p}\rangle}{{\rm d}t}=-\langle V'(\hat{x})\rangle
\end{equation}
which for a non-quadratic potential is not a simple expectation value of
$\hat{x}$. One can formally expand 
\begin{eqnarray}
  \langle V'(\hat{x})\rangle&=& \langle V'(\langle\hat{x}\rangle+
  (\hat{x}-\langle\hat{x}\rangle))\rangle\nonumber\\
 &=& 
  V'(\langle\hat{x}\rangle)+ \frac{1}{2} V'''(\langle\hat{x}\rangle) (\Delta
  x)^2+\cdots  \label{V}
\end{eqnarray}
with the fluctuation $\Delta
x=\sqrt{\langle(\hat{x}-\langle\hat{x}\rangle)^2\rangle}$ and additional terms
that contain higher moments
$\langle(\hat{x}-\langle\hat{x}\rangle)^n\rangle$. 

Since moments of a quantum-mechanical state are degrees of freedom independent
of expectation values, the system of equations (\ref{x}) and (\ref{p}) is not
closed. However, by the same procedure, computing expectation values of
commutators as in (\ref{x}), one can derive new evolution equations for
$\Delta x$ and all other moments. In a semiclassical expansion, only finitely
many equations need be taken into account at any fixed order, making the
system manageable. Moreover, one can combine this expansion with an adiabatic
one in which moments are assumed to vary more slowly than expectation
values. With these expansions, as shown in \cite{EffAc}, one reproduces the
usual low-energy effective action \cite{EffAcQM} for anharmonic
systems. Higher orders in the adiabatic expansion provide higher-derivative
terms \cite{HigherTime}.

The procedure sketched is the right basis to derive effective actions for
canonical quantum systems. However, an application to quantum gravity requires
additional extensions, most importantly one to include constraints or gauge
properties, and to address the problem of time in the absence of an absolute
evolution parameter such as the one used in (\ref{x}). Also this extension is
available at the canonical level \cite{EffCons,EffConsRel,EffConsComp} and
allows one to solve the problem of time, at least semiclassically
\cite{EffTime,EffTimeLong,EffTimeCosmo}. One could apply these techniques to a
quantum version of the algebra (\ref{DD})--(\ref{HH}), whose symmetry
generators in the gauge theory of gravity are constraints. However, a last
ingredient required for a successful implementation is still being developed:
a generalization of canonical quantum mechanical effective methods to quantum
field theories.

At present, it is therefore difficult to include quantum
back-reaction in the algebra (\ref{HH}) to see how it could be corrected. One
expects higher time derivatives to arise from these corrections, as has been
established for quantum-mechanical systems \cite{HigherTime}, and therefore
quantum back-reaction contributes to higher-curvature terms. If the
corrections are purely higher curvature, they do not change the
hypersurface-deformation algebra. But common low-energy arguments stating that
effective quantum back-reaction in gravity is of higher-curvature form assume
that there is a Poincar\'e-invariant vacuum state to be expanded around. In
non-perturbative quantum gravity, especially one with a discrete spatial
structure such as loop quantum gravity, it is unlikely that there is any
exactly Poincar\'e-invariant state. Standard arguments for higher-curvature
effective actions then break down, and in addition to curvature invariants
there may well be other terms that modify the algebra (\ref{HH}), reflecting
new space-time structures in the presence of discreteness.

It is possible for quantum back-reaction to modify (\ref{HH}) and compete with
the quantum-geometry corrections of loop quantum gravity. Inverse-triad
corrections do not refer directly to the density or curvature scale and are
therefore safe from competition by quantum back-reaction; they can be
discussed separately and bounded observationally. But holonomy corrections
always compete with higher-curvature terms.

Any holonomy modification of (\ref{HH}) could, in principle, be undone by
modifications due to quantum back-reaction. However, a more detailed look at
how quantum back-reaction in constrained systems arises shows that whatever
modification may be present, it cannot remove all possible modifications by
holonomy corrections. The reason for this is the form of degrees of
freedom. Holonomy corrections change even the dependence of Hamiltonian
(constraints) on expectation values; they are modifications of the classical
dynamics motivated by quantum geometry. Quantum back-reaction gives rise to
corrections that depend on moments of a state, as in (\ref{V}). If one
computes Poisson brackets of constraints corrected by quantum back-reaction,
all correction terms will still depend on moments after taking the Poisson
bracket: Moments are based on polynomial expressions in $x$ and $p$, at least
of second degree, and the Poisson bracket of two polynomials of degree at
least two is always a polynomial of degree at least two. Quantum corrections
of (\ref{HH}) due to quantum back-reaction will therefore contain moments,
while those due to holonomy corrections do not.

Quantum back-reaction can cancel holonomy modifications only for special
states in which moments are strictly related to expectation values in a
specific way. Generically, these corrections, even though their magnitudes are
similar, provide different terms that do not cancel each other. Even in the
absence of consistent versions of (\ref{HH}) that include quantum
back-reaction, it remains meaningful to study holonomy corrections and use
their implications for quantum space-time structure, in the same spirit as
already discussed for inverse-triad corrections.

\subsection{Holonomy corrections and signature change}

Holonomy corrections give rise to a difference equation (\ref{Diff}) for the
wave function, and imply strong modifications at high density. However,
quantum back-reaction and higher-curvature corrections are both significant in
the same regime, and therefore the high-curvature behavior of loop quantum
cosmology remains uncertain.  It is, however, clear that holonomy corrections
imply drastic effects on quantum space-time. The hypersurface-deformation
algebra with holonomy corrections is not completely known, but all cases that
have been computed so far give the same structure
\cite{JR,ThreeDeform,ScalarHol}:
\begin{eqnarray}
 [S(\vec{w}_1),S(\vec{w}_2)]&=& S((\vec{w}_2\cdot\vec{\nabla})\vec{w}_1-
 (\vec{w}_1\cdot\vec{\nabla})\vec{w}_2) \label{DDbeta}\\
{} [T(N),S(\vec{w})] &=& T(\vec{w}\cdot\vec{\nabla}N)\\
{} [T(N_1),T(N_2)] &=&
S(\beta(N_1\vec{\nabla}N_2-N_2\vec{\nabla}N_1)) \label{HHbeta} 
\end{eqnarray}
with a correction function $\beta$ that satisfies $\beta<0$ at high density
(the putative ``bounce'' of simple models). With modifications as in
(\ref{ModFried}), we have $\beta=\cos(2\ell H)$ in terms of the Hubble
parameter. Notice that inhomogeneities, although they must be present to have
a non-trivial derivation of the algebra (\ref{HHbeta}), are not the reason for
the modification by $\beta$. The reason is holonomy modifications, which
already appear for homogeneous background evolution. Inhomogeneity is merely
used to show non-trivial space-time effects, as the right-hand side of
(\ref{DDbeta})--(\ref{HHbeta}) vanishes identically for homogeneous $N$ and
$\vec{w}$.

\begin{figure}
  \includegraphics[height=.2\textheight]{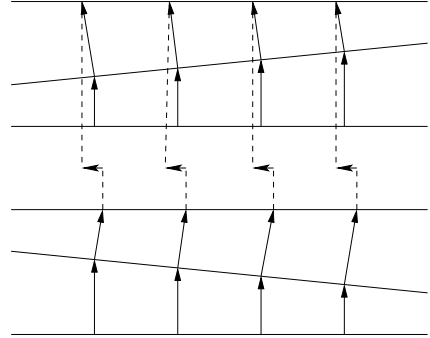}
  \caption{Holonomy corrections at high density imply a sign reversal in the
    correction function of (\ref{HHbeta}). The constructions of
    Fig.~\ref{HypDefLinMink} would indicate $\Delta x=-v\Delta t$, with
    reversed displacement arrows. A better geometrical explanation of the
    reversed sign and arrows is signature change: Redrawing
    Fig.~\ref{HypDefLinMink} with right angles according to Euclidean
    geometry, as shown here, implies the same reversal of spatial-displacement
    arrows as follows from a reversed sign in
    (\ref{HHbeta}). \label{HypDefLin1}}
\end{figure}

A negative $\beta$, for instance $\beta=-1$ at high density, means that
constructions such as those in Fig.~\ref{HypDefLinMink} lead to intransigent
motion with $\Delta x=-v\Delta t$, a displacement opposite to the velocity. A
geometrical interpretation is now more meaningful: A negative $\beta$ implies
that the space-time signature becomes Euclidean \cite{Action}, as can be seen
by comparing Fig.~\ref{HypDefLinMink} with the Euclidean version
Fig.~\ref{HypDefLin1}. Indeed, if one computes cosmological perturbation
equations analogous to (\ref{u}) and (\ref{w}), as done in
\cite{ScalarHol,ScalarTensorHol}, the positive $\alpha^2$ is replaced by
$\beta$, giving rise to an elliptic differential equation when
$\beta<0$. (Holonomy corrections, so far, do not lead to different parameters
$\beta$ and $s(\beta)$ for the independent modes of cosmological perturbation
equations.) The same effect happens when inhomogeneity is treated
non-perturbatively in spherically symmetric models \cite{JR}, where it can be
shown to be largely insensitive to quantization ambiguities
\cite{Action}. With signature change at high density, the cosmological
scenario of loop quantum cosmology is not a bounce, but is reminiscent of the
Hartle--Hawking picture, now derived as a consequence of quantum space-time
structure in the presence of holonomy corrections.

\section{Multiverse?}

In view of these new results, we must revise our multiverse scenario sketched
in the beginning of this article, based on cosmological bounces. (See also
\cite{Silence}.) We observed that inhomogeneous collapse combined with a
transition to expansion (a ``bounce'') may lead to causally disconnected
regions, expanding within a larger multiverse. Inhomogeneity of this type is
extremely hard to control with present-day non-perturbative quantum gravity,
but good effective methods are now available to help us understand the
relevant space-time structure.  Loop quantum gravity, it turns out, implies
radical modifications at Planckian densities, with a quantum version of
4-dimensional Euclidean space instead of space-time.

In Euclidean space, initial-value problems are ill-posed and there is no
propagation of structure from collapse to expansion, as assumed in bounce
models. Expanding patches that may result are causally disconnected not just
from their surrounding space-time, but also from their predecessor which was
collapsing.  Instead of a bounce, loop quantum cosmology, once inhomogeneity
is taken into account consistently, gives rise to a non-singular beginning of
the expanding Lorentzian phase we can observe. The transition from Euclidean
to Lorentzian signature, when $\beta=0$, is a natural place to pose initial
conditions, for instance for an inflaton state. These initial values are
unaware of what happened in the collapse phase, so that the picture of dense
collapsing patches bouncing first is not realized.

The new signature-change of loop quantum cosmology shares with bounces the
combination of collapse with expansion, but the collapse phase does not
deterministically affect the expansion phase. As a consequence, there is no
entropy problem because no complete information is transmitted through high
densities. And yet, the model is non-singular \cite{NoSing}.

One may still view the possible collection of expanding universes within one
space(-time), combining Euclidean and Lorentzian pieces, as a
multiverse. However, any causal contact realized is even weaker than what is
usually possible in multiverses, and it may seem more appropriate to talk of
separate universes instead of one however connected larger structure. Each of
these expanding patches has its own beginning when space-time emerges by
signature change from 4-dimensional space, giving it a clear status as a
universe of its own.

\section{Acknowledgements}

I am grateful to Mariusz Dabrowski and the organizers of the conference
Multicosmofun '12 for their invitation to give a talk on which this article is
based.  This work was supported in part by NSF grant PHY0748336.


\begin{thebibliography}{58}
\expandafter\ifx\csname natexlab\endcsname\relax\def\natexlab#1{#1}\fi
\providecommand{\enquote}[1]{``#1''}
\expandafter\ifx\csname url\endcsname\relax
  \def\url#1{\texttt{#1}}\fi
\expandafter\ifx\csname urlprefix\endcsname\relax\def\urlprefix{URL }\fi
\providecommand{\eprint}[2][]{\url{#2}}

\bibitem[Bojowald(2008)]{LivRev}
M.~Bojowald, \emph{Living Rev.\ Relativity} \textbf{11}, 4 (2008), {\tt
  http://www.livingreviews.org/lrr-2008-4}, \eprint{gr-qc/0601085}.

\bibitem[Bojowald(2011{\natexlab{a}})]{Springer}
M.~Bojowald, \emph{Quantum Cosmology: A Fundamental Theory of the Universe},
  Springer, New York, 2011{\natexlab{a}}.

\bibitem[Bojowald(2001{\natexlab{a}})]{Sing}
M.~Bojowald, \emph{Phys.\ Rev.\ Lett.} \textbf{86}, 5227--5230
  (2001{\natexlab{a}}), \eprint{gr-qc/0102069}.

\bibitem[Bojowald(2002)]{IsoCosmo}
M.~Bojowald, \emph{Class.\ Quantum Grav.} \textbf{19}, 2717--2741 (2002),
  \eprint{gr-qc/0202077}.

\bibitem[Bojowald(2001{\natexlab{b}})]{SemiClass}
M.~Bojowald, \emph{Class.\ Quantum Grav.} \textbf{18}, L109--L116
  (2001{\natexlab{b}}), \eprint{gr-qc/0105113}.

\bibitem[Ashtekar et~al.(2006)]{APS}
A.~Ashtekar, T.~Pawlowski, and P.~Singh, \emph{Phys.\ Rev.\ D} \textbf{73},
  124038 (2006), \eprint{gr-qc/0604013}.

\bibitem[Bojowald(2007)]{BouncePert}
M.~Bojowald, \emph{Phys.\ Rev.\ D} \textbf{75}, 081301(R) (2007),
  \eprint{gr-qc/0608100}.

\bibitem[Tolman(1934)]{Tolman}
R.~C. Tolman, \emph{Relativity, Thermodynamics and Cosmology}, Clarendon Press,
  Oxford, 1934.

\bibitem[Ashtekar and Bojowald(2006)]{BHInt}
A.~Ashtekar, and M.~Bojowald, \emph{Class.\ Quantum Grav.} \textbf{23},
  391--411 (2006), \eprint{gr-qc/0509075}.

\bibitem[Bojowald(2011{\natexlab{b}})]{Once}
M.~Bojowald, \emph{Once Before Time: A Whole Story of the Universe}, Knopf, New
  York, 2011{\natexlab{b}}.

\bibitem[Dirac(1958)]{DiracHamGR}
P.~A.~M. Dirac, \emph{Proc.\ Roy.\ Soc.\ A} \textbf{246}, 333--343 (1958).

\bibitem[Bojowald and Paily(2012{\natexlab{a}})]{Action}
M.~Bojowald, and G.~M. Paily, \emph{Phys.\ Rev.\ D} \textbf{86}, 104018
  (2012{\natexlab{a}}), \eprint{arXiv:1112.1899}.

\bibitem[Hojman et~al.(1976)]{Regained}
S.~A. Hojman, K.~Kucha\v{r}, and C.~Teitelboim, \emph{Ann.\ Phys.\ (New York)}
  \textbf{96}, 88--135 (1976).

\bibitem[Kucha\v{r}(1974)]{LagrangianRegained}
K.~V. Kucha\v{r}, \emph{J.\ Math.\ Phys.} \textbf{15}, 708--715 (1974).

\bibitem[Bojowald et~al.(2008)]{ConstraintAlgebra}
M.~Bojowald, G.~Hossain, M.~Kagan, and S.~Shankaranarayanan, \emph{Phys.\ Rev.\
  D} \textbf{78}, 063547 (2008), \eprint{arXiv:0806.3929}.

\bibitem[Bojowald et~al.(2009{\natexlab{a}})]{LTBII}
M.~Bojowald, J.~D. Reyes, and R.~Tibrewala, \emph{Phys.\ Rev.\ D} \textbf{80},
  084002 (2009{\natexlab{a}}), \eprint{arXiv:0906.4767}.

\bibitem[Reyes(2009)]{JR}
J.~D. Reyes, \emph{Spherically Symmetric Loop Quantum Gravity: Connections to
  2-Dimensional Models and Applications to Gravitational Collapse}, Ph.D.
  thesis, The Pennsylvania State University (2009).

\bibitem[Perez and Pranzetti(2010)]{ThreeDeform}
A.~Perez, and D.~Pranzetti, \emph{Class.\ Quantum Grav.} \textbf{27}, 145009
  (2010), \eprint{arXiv:1001.3292}.

\bibitem[Cailleteau et~al.(2012{\natexlab{a}})]{ScalarHol}
T.~Cailleteau, J.~Mielczarek, A.~Barrau, and J.~Grain, \emph{Class.\ Quant.\
  Grav.} \textbf{29}, 095010 (2012{\natexlab{a}}), \eprint{arXiv:1111.3535}.

\bibitem[Kreienbuehl et~al.(2012)]{ModCollapse}
A.~Kreienbuehl, V.~Husain, and S.~S. Seahra, \emph{Class.\ Quantum Grav.}
  \textbf{29}, 095008 (2012), \eprint{arXiv:1011.2381}.

\bibitem[Henderson et~al.(????{\natexlab{a}})]{TwoPlusOneDef}
A.~Henderson, A.~Laddha, and C.~Tomlin  (2012{\natexlab{a}}),
  \eprint{arXiv:1204.0211}.

\bibitem[Henderson et~al.(????{\natexlab{b}})]{TwoPlusOneDef2}
A.~Henderson, A.~Laddha, and C.~Tomlin  (2012{\natexlab{b}}),
  \eprint{arXiv:1210.3960}.

\bibitem[Tomlin and Varadarajan(????)]{AnoFreeWeak}
C.~Tomlin, and M.~Varadarajan  (2012), \eprint{arXiv:1210.6869}.

\bibitem[Varadarajan(????)]{AnoFreeWeakDiff}
M.~Varadarajan  (2012), \eprint{arXiv:1210.6877}.

\bibitem[Bojowald(2012)]{ReviewEff}
M.~Bojowald, \emph{Class.\ Quantum Grav.} \textbf{29}, 213001 (2012),
  \eprint{arXiv:1209.3403}.

\bibitem[Rovelli(2004)]{Rov}
C.~Rovelli, \emph{Quantum Gravity}, Cambridge University Press, Cambridge, UK,
  2004.

\bibitem[Thiemann(2007)]{ThomasRev}
T.~Thiemann, \emph{Introduction to Modern Canonical Quantum General
  Relativity}, Cambridge University Press, Cambridge, UK, 2007,
  \eprint{gr-qc/0110034}.

\bibitem[Rovelli and Smolin(1995)]{RS:Spinnet}
C.~Rovelli, and L.~Smolin, \emph{Phys.\ Rev.\ D} \textbf{52}, 5743--5759
  (1995).

\bibitem[Rovelli and Smolin(1994)]{RS:Ham}
C.~Rovelli, and L.~Smolin, \emph{Phys.\ Rev.\ Lett.} \textbf{72}, 446--449
  (1994), \eprint{gr-qc/9308002}.

\bibitem[Thiemann(1998{\natexlab{a}})]{QSDI}
T.~Thiemann, \emph{Class.\ Quantum Grav.} \textbf{15}, 839--873
  (1998{\natexlab{a}}), \eprint{gr-qc/9606089}.

\bibitem[Bojowald and Skirzewski(2006)]{EffAc}
M.~Bojowald, and A.~Skirzewski, \emph{Rev.\ Math.\ Phys.} \textbf{18}, 713--745
  (2006), \eprint{math-ph/0511043}.

\bibitem[Bojowald and Skirzewski(2007)]{Karpacz}
M.~Bojowald, and A.~Skirzewski, \emph{Int.\ J.\ Geom.\ Meth.\ Mod.\ Phys.}
  \textbf{4}, 25--52 (2007), proceedings of ``Current Mathematical Topics in
  Gravitation and Cosmology'' (42nd Karpacz Winter School of Theoretical
  Physics), Ed.\ Borowiec, A.\ and Francaviglia, M., \eprint{hep-th/0606232}.

\bibitem[Bojowald et~al.(2012)]{HigherTime}
M.~Bojowald, S.~Brahma, and E.~Nelson, \emph{Phys.\ Rev.\ D} \textbf{86},
  105004 (2012), \eprint{arXiv:1208.1242}.

\bibitem[Thiemann(1998{\natexlab{b}})]{QSDV}
T.~Thiemann, \emph{Class.\ Quantum Grav.} \textbf{15}, 1281--1314
  (1998{\natexlab{b}}), \eprint{gr-qc/9705019}.

\bibitem[Bojowald(2001{\natexlab{c}})]{InvScale}
M.~Bojowald, \emph{Phys.\ Rev.\ D} \textbf{64}, 084018 (2001{\natexlab{c}}),
  \eprint{gr-qc/0105067}.

\bibitem[Bojowald et~al.(2007)]{QuantCorrPert}
M.~Bojowald, H.~Hern\'andez, M.~Kagan, and A.~Skirzewski, \emph{Phys.\ Rev.\ D}
  \textbf{75}, 064022 (2007), \eprint{gr-qc/0611112}.

\bibitem[Bojowald et~al.(2001)]{InflTest}
M.~Bojowald, G.~Calcagni, and S.~Tsujikawa, \emph{JCAP} \textbf{11}, 046
  (2001), \eprint{arXiv:1107.1540}.

\bibitem[Bojowald and Hossain(2008)]{Tensor}
M.~Bojowald, and G.~Hossain, \emph{Phys.\ Rev.\ D} \textbf{77}, 023508 (2008),
  \eprint{arXiv:0709.2365}.

\bibitem[Bojowald and Calcagni(2011)]{LoopMuk}
M.~Bojowald, and G.~Calcagni, \emph{JCAP} \textbf{1103}, 032 (2011),
  \eprint{arXiv:1011.2779}.

\bibitem[Bojowald et~al.(2011{\natexlab{a}})]{InflConsist}
M.~Bojowald, G.~Calcagni, and S.~Tsujikawa, \emph{Phys.\ Rev.\ Lett.}
  \textbf{107}, 211302 (2011{\natexlab{a}}), \eprint{arXiv:1101.5391}.

\bibitem[Connes(1996)]{Connes}
A.~Connes, \emph{C.R.\ Acad.\ Sci.\ Paris} \textbf{323}, 1231--1235 (1996).

\bibitem[Doplicher et~al.(1995)]{NonCommST}
S.~Doplicher, K.~Fredenhagen, and J.~E. Roberts, \emph{Commun.\ Math.\ Phys.}
  \textbf{172}, 187--220 (1995), \eprint{hep-th/0303037}.

\bibitem[Calcagni(2010)]{Fractional}
G.~Calcagni, \emph{Phys.\ Rev.\ Lett.} \textbf{104}, 251301 (2010),
  \eprint{arXiv:0912.3142}.

\bibitem[Bojowald(2010)]{CUP}
M.~Bojowald, \emph{Canonical Gravity and Applications: Cosmology, Black Holes,
  and Quantum Gravity}, Cambridge University Press, Cambridge, 2010.

\bibitem[Bojowald and Paily(????)]{DeformedRel}
M.~Bojowald, and G.~M. Paily  (2012), \eprint{arXiv:1212.4773}.

\bibitem[Burgess(2004)]{BurgessLivRev}
C.~P. Burgess, \emph{Living Rev.\ Relativity} \textbf{7} (2004),
  http://www.livingreviews.org/lrr-2004-5, \eprint{gr-qc/0311082}.

\bibitem[Donoghue(1994)]{EffectiveGR}
J.~F. Donoghue, \emph{Phys.\ Rev.\ D} \textbf{50}, 3874--3888 (1994),
  \eprint{gr-qc/9405057}.

\bibitem[Deruelle et~al.(2009)]{HigherCurvHam}
N.~Deruelle, M.~Sasaki, Y.~Sendouda, and D.~Yamauchi, \emph{Prog.\ Theor.\
  Phys.} \textbf{123}, 169--185 (2009), \eprint{arXiv:0908.0679}.

\bibitem[Cametti et~al.(2000)]{EffAcQM}
F.~Cametti, G.~Jona-Lasinio, C.~Presilla, and F.~Toninelli, \enquote{Comparison
  between quantum and classical dynamics in the effective action formalism,} in
  \emph{Proceedings of the International School of Physics ``Enrico Fermi'',
  Course CXLIII}, IOS Press, Amsterdam, 2000, pp. 431--448,
  \eprint{quant-ph/9910065}.

\bibitem[Bojowald et~al.(2009{\natexlab{b}})]{EffCons}
M.~Bojowald, B.~Sandh\"ofer, A.~Skirzewski, and A.~Tsobanjan, \emph{Rev.\
  Math.\ Phys.} \textbf{21}, 111--154 (2009{\natexlab{b}}),
  \eprint{arXiv:0804.3365}.

\bibitem[Bojowald and Tsobanjan(2009)]{EffConsRel}
M.~Bojowald, and A.~Tsobanjan, \emph{Phys.\ Rev.\ D} \textbf{80}, 125008
  (2009), \eprint{arXiv:0906.1772}.

\bibitem[Bojowald and Tsobanjan(2010)]{EffConsComp}
M.~Bojowald, and A.~Tsobanjan, \emph{Class.\ Quantum Grav.} \textbf{27}, 145004
  (2010), \eprint{arXiv:0911.4950}.

\bibitem[Bojowald et~al.(2011{\natexlab{b}})]{EffTime}
M.~Bojowald, P.~A. H\"ohn, and A.~Tsobanjan, \emph{Class.\ Quantum Grav.}
  \textbf{28}, 035006 (2011{\natexlab{b}}), \eprint{arXiv:1009.5953}.

\bibitem[Bojowald et~al.(2011{\natexlab{c}})]{EffTimeLong}
M.~Bojowald, P.~A. H\"ohn, and A.~Tsobanjan, \emph{Phys.\ Rev.\ D} \textbf{83},
  125023 (2011{\natexlab{c}}), \eprint{arXiv:1011.3040}.

\bibitem[H\"ohn et~al.(2012)]{EffTimeCosmo}
P.~A. H\"ohn, E.~Kubalova, and A.~Tsobanjan, \emph{Phys.\ Rev.\ D} \textbf{86},
  065014 (2012), \eprint{arXiv:1111.5193}.

\bibitem[Cailleteau et~al.(2012{\natexlab{b}})]{ScalarTensorHol}
T.~Cailleteau, A.~Barrau, J.~Grain, and F.~Vidotto, \emph{Phys.\ Rev.\ D}
  \textbf{86}, 087301 (2012{\natexlab{b}}), \eprint{arXiv:1206.6736}.

\bibitem[Mielczarek(????)]{Silence}
J.~Mielczarek  (2012), \eprint{arXiv:1212.3527}.

\bibitem[Bojowald and Paily(2012{\natexlab{b}})]{NoSing}
M.~Bojowald, and G.~M. Paily, \emph{Class.\ Quantum Grav.} \textbf{29}, 242002
  (2012{\natexlab{b}}), \eprint{arXiv:1206.5765}.

\end{thebibliography}

\end{document}